# Energy and GHG saving potentials of sufficiency measures - a synthesis for Germany

Carina Zell-Ziegler[1,2,a], Célia Burghardt[3], Kaya Dünzen[4], David Schöpf[5], Jenny Kurwan[5], Benjamin Best[6], Mirko Schäfer[3], Frauke Wiese[7]

[1] Oeko-Institut, Borkumstr. 2, 13189 Berlin, Germany
[2] Technische Universität Berlin, Straße des 17. Juni 135, 10623 Berlin, Germany
[3] INATECH, University of Freiburg, Emmy-Noether-Str. 2, 79110 Freiburg, Germany
[4] Oeko-Institut, Merzhauser Straße 173, 79100 Freiburg, Germany
[5] Wuppertal Institute for Climate, Environment and Energy, Döppersberg 19, 42103 Wuppertal, Germany
[6] City of Bonn, Climate Neutral Bonn 2035 Program Office, Bonn, Germany
[7] Europa-Universität Flensburg, Auf dem Campus 1, Flensburg 24941, Germany

[a] Corresponding author: c.zell-ziegler@oeko.de

## Abstract

The sixth assessment report of the Intergovernmental Panel on Climate Change (IPCC) emphasises the potential of demand-side measures, such as sufficiency, for mitigating climate change. Although quantified potentials of various sufficiency measures exist, policy advisors and energy modellers criticise the lack of findability and comparability of the relevant data. Due to the high level of heterogeneity in units, reference points and calculation methods, the data cannot be used to summarise sufficiency potentials at a national level. Consequently, this paper aims to identify, structure, harmonise and synthesise existing data in order to determine the greatest saving potentials per sector and highlight data gaps. Based on a systematic literature review, we have created a curated open-source database containing over 300 quantified energy and greenhouse gas (GHG) saving potentials for Germany, which could be used as a blueprint for other such data collections. Most quantifications were available for the building sector, particularly for appliances. The highest total energy and GHG emission savings in Germany were identified in measures that reduce per capita living space, with saving potentials of -150 Terawatt-hours per year (TWh/a) and -118 Million Tons of Carbon Dioxide Equivalents per year (Mt $CO_2$eq./a; this measure also includes lower heating temperatures). This synthesis can help modellers to better account for sufficiency potentials in scenarios, and help policymakers to understand the saving potentials of sufficiency. We encourage researchers to quantify more energy and GHG saving potentials in order to fill the identified gaps and to use the proposed synthesis structure.

# 1.  Introduction

## 1.1.  The role and status of energy sufficiency (policy)

As the climate crisis intensifies and efforts to combat climate change need to be increased, demand-side options and particularly those that go beyond green growth have become part of discussions (Akenji 2014; Lorek and Spangenberg 2014; Wachsmuth and Duscha 2019; IPCC 2022; Creutzig et al. 2024; Sugiyama et al. 2024). One of the discussed concepts for implementation is energy sufficiency (Darby and Fawcett 2018; Sandberg 2021; Jungell-Michelsson and Heikkurinen 2022; Lage 2022).

The Intergovernmental Panel on Climate Change (IPCC) acknowledged for the first time in 2022 that sufficiency is a key strategy for climate change mitigation and described sufficiency policies as "a set of measures and daily practices that avoid demand for energy, materials, land, and water while delivering human well-being for all within planetary boundaries" (IPCC 2022, p. 35). This dual focus on individual behaviours and systemic policies underscores the comprehensive scope of sufficiency as a sustainability strategy.

While there are various interpretations of energy sufficiency, all converge on a shared vision of reducing consumption and production to achieve environmental sustainability. This paper adopts the understanding and definition proposed by Lage et al. (2023) as "a strategy for reducing, in absolute terms, the consumption and production of end-use products and services through changes in social practices in order to comply with environmental sustainability while ensuring an adequate social foundation for all people". Table 2 in their paper presents exemplary sufficiency measures per sector. The emphasis on the social and equity aspect of energy sufficiency (social foundation), which is also found in the IPCC definition, is an important addition to the discussion on climate change mitigation efforts towards a just transition. It aligns with current research on well-being: demand-side options have been shown to be consistent with high levels of well-being (Creutzig et al. 2022a).

Another advantage of the energy sufficiency strategy is its multi-solving nature as it addresses different and interrelated objectives simultaneously. These include decarbonisation (chapter 1.2 looks at this in more detail), staying within planetary boundaries, limiting resource use, advancing sustainable development goals and promoting equity (Piketty and Chancel 2015; Roy et al. 2021; Wuppertal Institut 2023; SRU 2024). Unlike other approaches, which often involve trade-offs or the risk of undermining complementary goals, sufficiency seeks to minimise these by targeting multiple sustainability goals. For example, strategies such as the expansion of renewable energy or negative emission technologies may inadvertently lead to increased resource demand or inequalities in access to these resources because of high systemic costs. Reducing energy demand by means of sufficiency measures reduces the pressure to tap these options at unsustainable rates (Wiese et al. 2022, 2024a).

Researchers also shed light on different further aspects of sufficiency, its role within climate change mitigation and its governance: Sandberg (2021) provides nuance by identifying four types of consumption changes inherent to sufficiency, namely absolute reduction, modal shift, product longevity and sharing practices. These aspects emphasise the multifactorial nature of sufficiency, which operates not only at the individual level but also through broader political and systemic changes. Lage (2022) describes various notions of sufficiency and defines three different approaches to realising a transition towards sufficiency: a bottom-up-approach, a policy-making-approach and a social-movement-approach. Dablander et al. (2025) describe actions that different actors such as policymakers, businesses, NGOs, research, media / arts and civil society can take and are already taking towards sufficiency. Zell-Ziegler et al. (2025) present a collection of more than 350 policy instruments for different sectors, showing how sufficiency can be put into practice. This collection also includes some implementation examples. All of these authors emphasise the importance of sufficiency governance and policy instruments that create good framework conditions for the shift in social practices and cultural norms. However, as research also confirms, sufficiency as a policy instrument is still underrepresented in the current policy mix for decarbonising the economy and mitigating climate change (Zell-Ziegler et al. 2021; Lage et al. 2023).

## 1.2. Sufficiency potentials in the existing literature

In this article, we focus on the saving potentials of energy sufficiency expressed as greenhouse gas (GHG) emission and energy savings. There is a range of studies (peer-reviewed and grey) that has explored these saving potentials for different geographic scopes: On the global scale, the IPCC (2022) summarises the existing research and highlights a significant emission reduction potential of all demand-side strategies, ranging from 40-70 % in the end-use sectors buildings, land transport, and food (Creutzig et al. 2022b, p. 505). In another paper, Creutzig et al. (2022a) even set the potential at a 40-80 % reduction in end-use sector emissions brought about by socio-behavioural, infrastructural and technological options. In their low-energy demand scenario, Grubler et al. (2018) predict a 40 % overall reduction in final energy consumption by 2050 compared to 2018 levels, with the Global North achieving a reduction of almost 55 %. This would be achieved through, among other measures, the large-scale replacement of carbon-intensive steel with other materials and a halving of flight kilometres. Comparable scenarios which integrate lifestyle changes and sufficiency on the global level are provided, for example, in van Den Berg et al. (2024), which state a 39 % emission reduction in 2050 in the Global North in the transport and residential sectors compared to a baseline scenario.

At the regional level, the CLEVER scenario (Collaborative Low Energy Vision for the European Region) outlines an ambitious mitigation path. It specifies that the final energy demand in 2050 will more than halve, with at least 40 % of the reduction ascribed to sufficiency (Bourgeois et al. 2023; Wiese et al. 2024a). Another regional scenario orientated to the demand side is provided by Costa et al. (2021) for the EU+2 (the UK and Switzerland). They model a 43 % reduction in GHG emissions in the scenario with lifestyle changes compared to a baseline scenario. More than 20 % of this reduction can be attributed to behavioural change/sufficiency.

At the next level down, there are also a number of studies conducted at the country level. In the UK, the low-energy demand scenario by Barrett et al. (2022) models final energy consumption cut by 52 % compared to business-as-usual projections. The bottom-up modelling scenario offered by Bauer and Sterner (2025) for Germany shows a 61 % reduction in final energy consumption in 2050 compared to 2019 in the scenario which combines technological options with lifestyle changes. More European and country-specific reduction potentials for final energy demand and energy service parameters are compared in an analysis provided by Wiese et al. (2024b).

In addition to differences in geographic scopes, studies on the saving potentials of sufficiency also differ in their thematic focus. Some studies focus on single sectors, some focus on a selection of consumption or lifestyle changes (if these changes include measures but not a specific way or policy instrument for achieving them, like reduction of living space or sharing economy, we call this *cluster*). Other studies analyse the potentials of single policy instruments or measures in detail. A few studies and their quantified saving potentials are presented here as examples: Gaspard et al. (2023) focus on the reduction of energy consumption in the buildings sector in France and model four scenarios, two of them with an explicit focus on sufficiency. The most ambitious one reduces energy consumption by 40 %, compared to the business-as-usual scenario in 2050. Based on a literature review, Ivanova et al. (2020) present GHG mitigation potentials for 61 consumption options, some of which involve sufficiency, in the food, transport and housing sector and for other consumption options, from a global perspective. The highest saving potential was identified for living car-free with a mean of more than 2 t $CO_2$-eq./capita/a. Vita et al. (2019) modelled different green consumption and sufficiency options for Europe using a consumption-based perspective, also for different sectors, and compared them. They similarly accorded a mobility measure the highest saving potential: only bike and walk with 26 % lower carbon emissions than the baseline. Kreye et al. (2022) look at one very specific case of home office. They calculate the emission savings during the COVID-19 pandemic for Germany (almost 4 Mt $CO_2$-eq./a) and for two possible scenarios with greater or fewer days of working from home (almost 10 Mt $CO_2$-eq./a for the ambitious case). Paar et al. (2023) performed an ex-post evaluation of cycling projects in German municipalities. The highest saving potential was identified for the city of Wuppertal with 3 kt $CO_2$-eq./a for the years from 2020 and projected until 2045.

We can see that there is a variety of existing quantifications of GHG and energy saving potentials from sufficiency or related concepts with very different geographic or thematic foci. Figure 1 shows this heterogeneity in the existing literature.

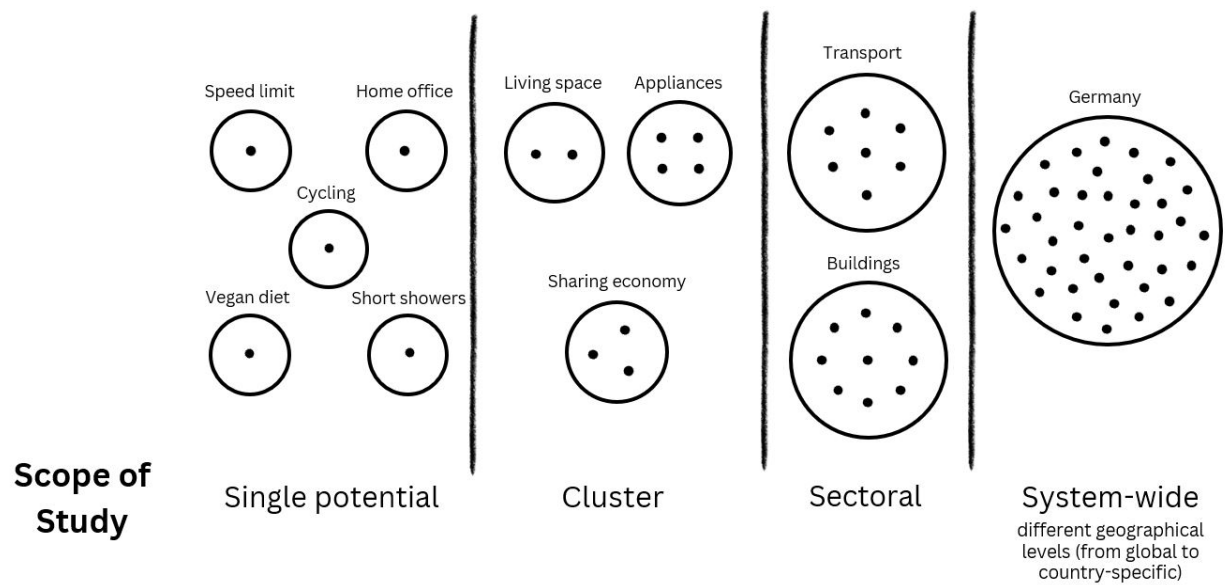

**Fig. 1** Heterogeneity of saving potentials in the existing literature
Legend: A circle stands for a study, a dot represents the quantified potential of a single measure.

## 1.3.  Research gap and aim of the study

We have described the relevant mitigation potentials of sufficiency options. However, we also mentioned that sufficiency does not currently play a significant role in policy. Energy models and climate mitigation scenarios are powerful tools that can help decision-makers design policy strategies and concrete instruments (Süsser et al. 2021). One reason for the lack of policy attention may therefore be the fact that sufficiency is so far under-represented in energy and climate mitigation modelling (Samadi et al. 2017; Golinucci et al. 2025).

In an expert workshop with German modellers and researchers, Zell-Ziegler and Förster (2018) found that modellers face a major hurdle in the form of a lack of available sufficiency saving potential data, which is scattered throughout the literature and difficult to find. McWilliams et al. (2025) add that another barrier to using existing data is the lack of harmonisation due to differences in units, assumptions and scales. Thus, as shown in the subchapter on existing potentials above, the research gap is not primarily about the lack of quantifications of sufficiency saving potentials, but rather its identification, structuring, harmonisation and synthesis. A subsequent research need is to analyse the existing data and highlight gaps for further exploration and research.

The study aims to address these gaps by creating a structured database of harmonised sufficiency saving potentials (energy and GHG emissions), which will be made publicly available. The focus of the study was set at the national and measure-specific levels ("single potential" and "cluster" in Figure 1) because decarbonisation targets and modelled scenarios are primarily developed at the national level, while the implementation of climate change policy happens at the measure level.

The article is structured as follows: Section 2 describes the methodology of the literature review, the creation and compilation of the database, the data harmonisation, the search for gaps and the presentation of results. The overall and sector-specific results are presented and discussed in section 3 followed by the data gap analysis. Finally, the main findings and recommendations (for future research) are presented in the conclusion (section 4).

# 2.  Material and Methods

For this analysis, we performed a systematic literature review on energy sufficiency saving potentials. Our national case study focuses on Germany. In parallel, we conceptualised a database structure to curate the data and make the potentials comparable. We then filled it with the results of the literature review. To identify data gaps, we compared the database entries with an extensive list of sufficiency policy instruments. All underlying materials and methods are described in the following sections.

## 2.1.  Systematic literature review

We conducted a systematic literature review (Tranfield et al. 2003) in the Web of Science Core Collection and in Google Scholar on May 25, 2023 in English and in German. From our research experience, we know that, at least in Germany, many applied research projects on sufficiency have not published their results in peer-reviewed journals, e. g. Fischer et al. (2016). Therefore, in addition to the scientific literature, we have also included grey literature in this review.

We used the search string presented in Figure 2 (English) and the corresponding search string in German, see Figure 8 in the Appendix.

$$\{energy\} \; AND \; \left\{ \{sufficiency\} \; OR \; \left\{ \begin{Bmatrix} \{reduc*\} \\ OR \\ \{absolute \; reduction\} \\ OR \\ \{sav*\} \\ OR \\ \{avoid\} \end{Bmatrix} \; AND \; \begin{Bmatrix} \{demand\} \\ OR \\ \{behaviou*\} \\ OR \\ \{lifestyle\} \\ OR \\ \{consumption\} \end{Bmatrix} \right\} \right\} \; AND \; \begin{Bmatrix} \{quanti*\} \\ OR \\ \{calculat*\} \\ OR \\ \{empirical\} \end{Bmatrix} \; AND \; \{Germany\} \; NOT \; \begin{Bmatrix} \{self-sufficiency\} \\ OR \\ \{flexib*\} \end{Bmatrix}$$

**Fig. 2** Search query

The definition of the search string followed an iterative process: keywords like *energy* and *sufficiency* and the geographic focus on Germany were set from the beginning. Then we tried to paraphrase sufficiency with a combination of different terms which we took from relevant papers such as Sandberg (2021), Ivanova et al. (2020), IPCC (2022), Vita et al. (2019) and we checked whether suitable articles could be found. Additional search terms were incorporated to address the quantitative nature of the articles, meaning that actual quantified saving potentials should be provided. Finally, terms were added to exclude articles related to self-sufficiency and flexibility as these frequently produced irrelevant results. We only included articles and documents published from 2013 onwards.

We complemented this systematic literature review with additional papers published up to May 2024. Therefore, we contacted the German and European communities on energy sufficiency via the mailinglists "Forschungsnetzwerk-Suffizienz"[1] (German research network on sufficiency) and "ENOUGH network"[2] in spring 2024 and screened the Zotero literature database of the ENOUGH network[3] using the keywords *German* and *Germany*. Additional papers known to the authors or identified through snowball sampling were included if they were absent from the collection after completing these steps. This led to 430 papers from the systematic literature search and 85 additional papers, see Figure 3.

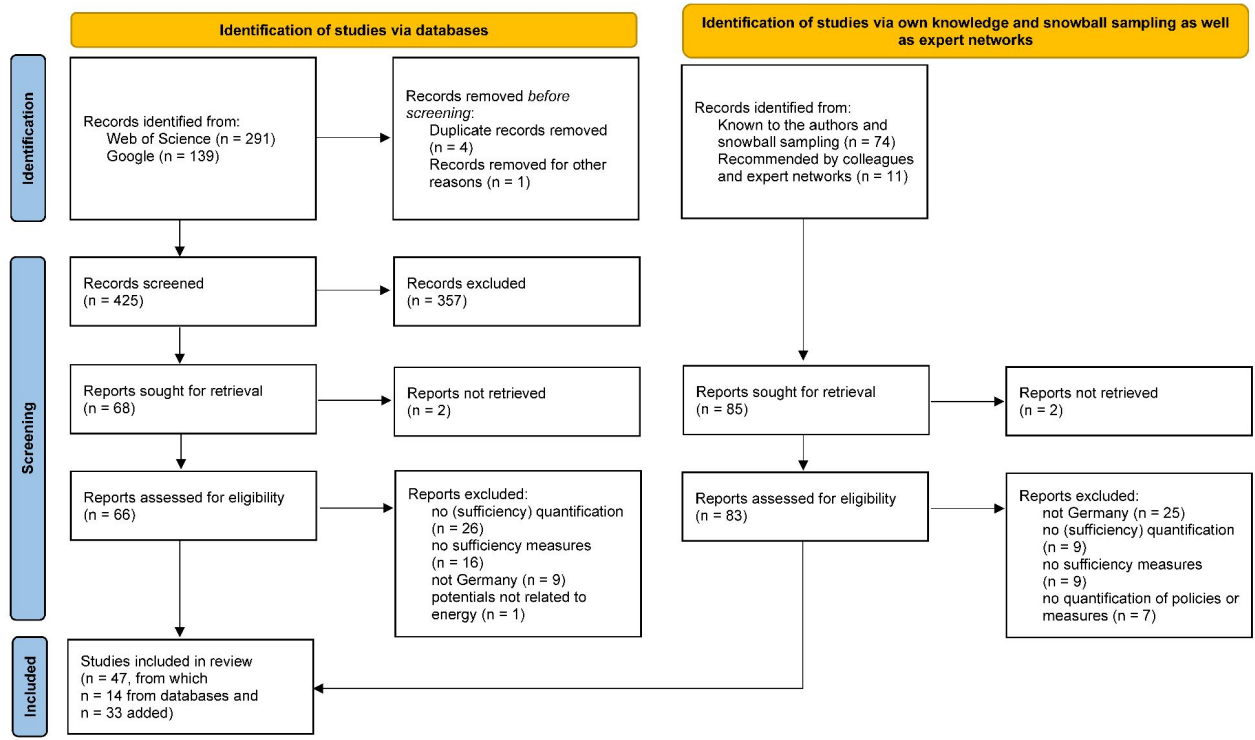

**Fig. 3** PRISMA flow diagram on the systematic literature review based on Page et al. (2021)

After removing duplicates and one paper published before 2013, we screened the abstracts of 425 papers. The majority (84 %) were excluded because they did not meet the search criteria, e. g. they did not focus on energy sufficiency but used the key terms in other contexts. Another common reason was that the article mentioned Germany in the text but did not focus on it.

This left 66 papers found in databases and 83 papers added by experts for a detailed screening. We clustered our decisions into *include* and used five exclusion criteria: *no (sufficiency)*

---

*quantification, not Germany, no sufficiency measures, no quantification of policies or measures* and *potentials not related to energy.* 102 reports (68 %) were excluded in this step. This resulted in 47 papers remaining for our analysis, 14 of which originated from the databases Web of Science and Google Scholar and 33 from the additions.

These screening steps were performed by the team of eight authors. Inter-coder validity (see, for example, Olson et al. (2016)) was ensured through regular coder meetings to discuss inclusion / exclusion criteria and other questions. If coders were unsure about the inclusion or exclusion of a study, a second coder re-evaluated the abstract and/or study.

## 2.2. Design of the database

There is no comparable data collection available, so we structured the *Sufficiency Potential Database* from scratch. Every energy or GHG saving measure fills one row. Besides the name of the measure and its energy and/or GHG saving potential, we added a lot of contextual information / metadata which is essential to ensure the understanding of the exact measure and its respective saving potential, thereby enhancing the usability of the data. As shown in chapter 1.2, the existing literature on potentials is very heterogeneous in terms of geographic and thematic scope. Additionally, the methods used, assumptions made, the years modelled etc. vary. Two types of contextual information on the potentials are important: 1) context information on the exact saving potential like the unit, the respective year to which it applies, a description of the measure and the relevant sector and 2) context information on how the saving potential was determined in the study, such as information on the type or quality (estimation, modelling result or ex-post evaluation) or the calculation method (territorial or consumption-based approach), which can explain why a potential is high or low.

The design of the *Sufficiency Potential Database* followed an iterative process. The first potentials from about ten studies were extracted for testing in early 2023. The differences found for these studies and potentials led to the prototype of the database. Following discussions among the author team and external experts, the columns were amended and several new columns were added. The idea and the prototype were presented at two scientific conferences (NFDI4Energy conference and eceee Summer Study, both in 2024) and discussed with modellers and energy efficiency and sufficiency experts to assure its usability for the target group. The final structure of the database with all metadata as columns can be found in the Supplementary Information.

To enhance comparability and visualisation of potentials, we clustered all database entries. The clustering was done sector-by-sector by screening all entries and deductively attributing clusters. We attributed a higher-level cluster (e. g. heat consumption) and a cluster (e. g. living space or heating behaviour) to each potential. Measures were categorised as *single* or *combined* based on the cluster. Entries linked to only one cluster were labeled as *single*, while 42 entries were classified as *combined* because they were attributed to two or more clusters (e. g. the reduction of motorised individual transport through more public transport and active mobility with the latter two being separate clusters). The *single or combined* column enables filtering, which allows for a focus on specific saving potentials, such as active mobility, without broader cluster overlaps.

There are two scopes of the database: the *wide scope* includes all saving potentials related to sufficiency. The *narrow scope* only includes entries classified as pure sufficiency without an overlap with other mitigation strategies and only those classified as *single* measures.

The database was set up as a spreadsheet and can be found as Supplementary Information.

## 2.3. Filling of the database and conversion steps

All authors of this study filled the database with saving potentials from the identified 47 studies - per measure either energy or GHG savings or both, depending on the availability. All of the entries to the database were reviewed by a second person to ensure the quality of the entry and to streamline the wording. The definition of clusters was done by one author and revised by two others.

The original values from the studies are provided in the sheet *DB_original_values*. To harmonise the entries and enhance comparability and usability, conversion steps were undertaken. The results can be found in the sheet *DB_converted*, the conversion factors in the sheet *Conversion factors*. The conversion included the following steps:

- Calculating the average if minimum and maximum values were given.
- Calculating yearly values if the saving potential was given as a cumulative value: division by number of years (under the simplified assumption of constant annual savings).
- Converting the values for GHG mitigation potentials which were not given in Mt $CO_2$-eq./a (only 2 % of entries remained which could not be converted due to units like potential per m²) to enable comparison of the entries in the same unit.
- Converting values for energy saving potentials which were not given in TWh/a (we have only 7 % of entries left which could not be converted due to units like savings per m²) to enable comparison of the entries in the same unit. For the conversion of fuel units in litre or weight, we used the calorific values of petrol and diesel provided by the Massachusetts Institute of Technology (MIT 2007).
- Converting the twelve saving potentials that were given either per individual or per household, using the actual numbers for Germany from the Federal Statistical Office.

The resulting curated database is the first of its kind and is intended as an openly accessible tool for modellers, policymakers and anyone involved in designing and implementing decarbonisation options. Although this version of the database has a national focus on Germany, it can serve as a blueprint for collections for other countries or be extended to other countries in the future. The structure of the database also provides guidance to researchers who quantify saving potentials by indicating what metadata are needed to effectively classify, compare and reuse their results.

## 2.4. Gap analysis

The *Sufficiency Potential Database* was analysed to identify gap*s* in the quantification of individual measures, e. g. to recommend foci to modellers for future analysis. As a reference, we used the *European Sufficiency Policy Database* (Zell-Ziegler et al. 2025), an extensive sufficiency policy database with more than 350 entries and a focus on Germany and Europe. Firstly, we compared the sector-based entries of both databases and then compared the clusters of the *Sufficiency Potential Database* to the measures and actions in the policy database. This was followed by a detailed matching of *Sufficiency Potential Database* entries to specific policy instruments in the *Sufficiency Policy Database* (see Supplementary Information, *DB_converted*, columns AQ and AR).

## 2.5. Presentation of results

We summarise the saving potentials from a sectoral perspective and create corresponding figures. In doing so, we combine various distinguishing features: We always show the time scale and as a second factor, either the type of potential or the method of calculation. If all potentials have the same characteristics for these two factors, we only show the time scale and indicate the characteristics for these factors in the figure legend.

We produce separate figures for energy and GHG savings. All figures only show the narrow scope of the sufficiency potentials (see definition in chapter 2.2). In the figures, only potentials with the unit we used for comparison (TWh/a or Mt $CO_2$-eq./a) are shown. Additionally, only potentials for the national level are shown; these are valid for Germany as a whole. The very low local potentials are not included in the figures. For the figures on energy savings, we compare the potentials with the current final energy demand of the respective sector, retrieved from AGEB (2024) so as to categorise the level of the reduction potential.

We use the higher-level clusters and clusters to structure the figures: the clusters are used as figure labels on the x-axis, the higher-level clusters are given after the cluster name in parentheses and structure the x-axis. The higher-level clusters are separated from each other optically by the vertical dashed lines. Comparisons of saving potentials make particular sense within higher-level clusters and within clusters. A comparison of different clusters needs to take the different scopes of the clusters into account (e. g. in the cases of *appliances*, it is difficult to compare the electricity savings achieved through sufficiency with broader measures such as remote working or natural carbon sequestration).

Which exact measures and potentials lie behind each dot in the figures can be looked up in the database, see Supplementary Information (*DB_converted*), together with all metadata for this data point.

# 3. Results and discussion

## 3.1. Overview of database entries

Based on the literature review, the *Sufficiency Potential Database* (see Supplementary Information) was filled with findings from 47 literature sources. It contains 303 entries, 209 of which have a narrow scope (see definition in chapter 2.2). The 94 potentials that go beyond the narrow scope are included because of their valuable demand-side data. Six studies with 27 database entries explicitly explore combined potentials; these illustrate, for example, the interplay of sufficiency and efficiency, which often cannot be meaningfully separated.

From the 47 literature sources, some contributed only a few entries, while others provided many more: For example, the German 2023 Projections Report by Harthan et al. (2023) contributed 36 entries. 60 % of the entries fall within the narrow scope of our database, primarily in the industry / production and land-use & food sectors. The 36 entries are based on 15 measures, each of which have saving potentials for two to three points in time. The study by Fischer et al. (2016) contributed 32 entries; it focuses entirely on pure sufficiency and thus falls within our narrow scope and covers all sectors with quite detailed measures (e. g. behavioural measures like *watch less TV*). The data points of different studies differ in quite a lot of characteristics, that is why the metadata we provide with each potential like base year and calculation method is very important. In the following, numbers, percentages and the figures relate only to the narrow scope if not otherwise stated.

The building sector accounts for the largest share of entries (41 %), with about a third focusing on appliances. This focus is based on the fact that there are few very detailed studies on sufficiency measures in this sector, resulting in a large number of (very detailed) entries in the database.

64 % of entries contain GHG savings, compared to 49 % with energy savings, 12 % of the entries provide saving potentials on both metrics. GHG saving potentials are helpful for discussions on the policy level. Modellers need energy savings to integrate the sufficiency potentials into their models or rather activity levels like heated floor area or travelled kilometres. The latter lies outside the scope of our analysis and can be drawn from other papers like Wiese et al. (2024b).
Beyond GHG and energy savings, about 20 % of the entries and 10 of the 47 studies included in this analysis provide additional metrics such as material savings, land use, costs, air quality, noise reduction, road accidents, toxicity, and acidification. These metrics are very helpful for emphasising the multi-solving nature of sufficiency, which is not just a decarbonisation strategy. It would be helpful if more studies provided such additional metrics.

The database includes a broad range of saving potentials, from negligible to over 100 TWh or Mt $CO_2$-eq per year. Very notable energy saving measures in this regard include lowering per capita living space from 47 to 30 m²/person with an energy saving potential of 150 TWh/a (Bierwirth and Thomas 2019) and the measure *car use instead of ownership (people with good public transport connections do without cars)* with a saving potential of 104 TWh/a (Fischer et al. 2016). Both potentials account for 15 % of the respective sector's current final energy demand. Regarding GHG savings, sufficiency measures in buildings such as *reduction of room temperature* and *optimising space use* (with the per capita living space reducing from 50 to 30 m²/person) show a substantial reduction potential; the combination of these measures enables savings of up to 118 Mt $CO_2$-eq./a (Zimmermann 2018).

Most of the database entries involve theoretical potentials (85 %), which are mainly derived from estimates and model simulations. This characteristic results in a high variability of saving potentials due to different underlying assumptions. In contrast, realised potentials (quantified on the basis of ex-post evaluations) account for only 4 % (nine entries), six of which are visualised in the sectoral figures in the next subchapter and the Appendix. These realised saving potentials show rather small energy or GHG savings which is to be expected when compared to, for example, theoretical potentials. Ex-ante evaluations of implemented measures were only available in one paper (Paar et al. 2023), which focused on local cycling projects. Reports on bottom-up, already implemented measures like this are of special interest to local or regional administrations who are considering the adoption of measures that have already been empirically evaluated. It would thus be desirable to have more (ex-ante as well as ex-post) evaluations like this. The database entries cover all temporal scales, predominantly in the medium term (from 2030 to 2039, 38 %), followed by entries on potentials valid for the years up to 2024 (labelled *past*, 33 %), long-term projections (post-2040, 24 %), and a few saving potentials for the short term (between 2025 and 2029, 4 %).

In the course of the gap analysis (comparison of the database to the *Sufficiency Policy Database*), we identified how many of the measures are backed by a policy instrument[4]. This is because in many studies, decarbonisation measures are proposed without clarifying how they can be realised (Thema 2024). The corresponding share in the database is one third (34 %). None of the measures backed by a policy instrument is among the top saving measures (see mentioning of some above).

## 3.2. Sectoral saving potentials

This chapter presents the sectoral saving potentials. Chapter 2.5 provides explanations of how the results are presented in the figures. The five cross-sectoral saving potentials are not discussed in the following; the corresponding figures can be found in the Appendix (Figure 14).

### 3.2.1. Building sector

Energy-saving quantifications for the building sector rely exclusively on a territorial (non-LCA (life cycle assessment)) approach and comprise theoretical potentials. The database includes 62 entries for energy reduction measures, predominantly in the clusters *living space*, *heating behaviour*, and *appliances*, see Figure 4.

For *appliances*, reduction potentials range from small-scale measures, such as optimising ventilators and TVs (below 1 TWh/a), to more impactful measures targeting larger appliances like dryers and refrigerators or addressing multiple appliances within buildings (up to 19.8 TWh/a, see Fischer et al. (2013)). In the *living space* cluster, reducing average living space from 46.6 m² to 30 m² per person yields the highest potential, saving 150 TWh/a (Bierwirth and Thomas 2019), which is equivalent to 15 % of the energy demand of the 2022 building sector in 2022. Less ambitious measures, such as reductions to 40 m² per person, achieve 25.5 TWh/a (Fischer et al. 2016), while those solely targeting specific household groups achieve only 0.02–0.2 TWh/a (Fischer et al. 2020). Measures targeting specific household groups (e. g. young couples, older people) also appear in the clusters *energy consultation* and *downsizing*. In the *heating behaviour* cluster, most measures focus on reducing room temperatures by 1–2°C for varying shares of buildings in Germany.

---

[4] We use the following definitions on measures and policy instruments in this article: "Sufficiency measures are activities that lead to a verifiable, measurable or assessable reduction in environmental impacts. In contrast, sufficiency instruments are government interventions with the aim to promote the implementation of sufficiency measures" (translated based on Zell-Ziegler and Förster 2018).

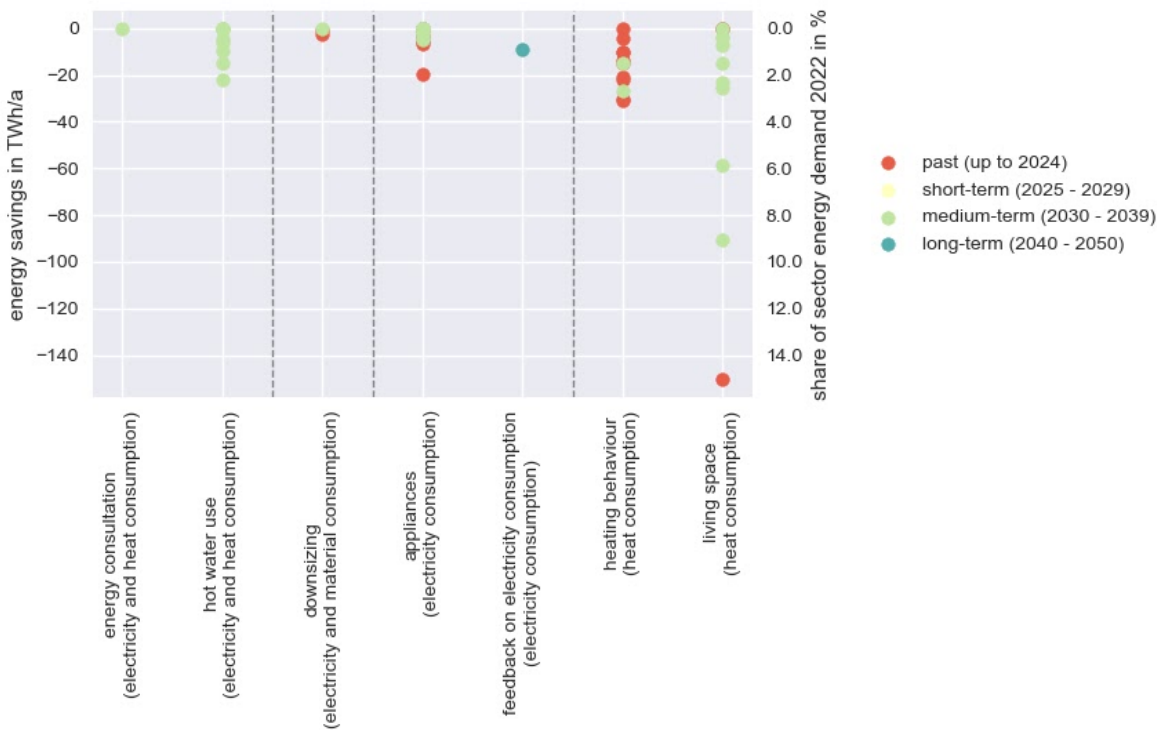

**Fig. 4** Energy saving potentials in the building sector. Characteristics of all potentials include territorial calculation method and type: theoretical. Final energy demand data for the year 2022 was taken from AGEB (2024).

The database includes GHG reduction potentials for 23 measures; LCA was used to calculate four of them and a territorial calculation approach was used to calculate the remainder, see Figure 9 in the Appendix. All comprise theoretical potentials, and most measures fall within the *living space* cluster.The top three reductions originate from a single source (Zimmermann 2018) and are attributed to immediate living space reductions (e. g. from 100 m² to 60 m² for two-person households, saving 52.2 Mt $CO_2$-eq./a), ambitious heating behaviour changes (e. g. maintaining 18°C room temperature with efficient ventilation, -51.3 Mt $CO_2$-eq./a), and halving electricity demand (-32.3 Mt $CO_2$-eq./a).

For the *living space* cluster, ambition levels vary, ranging from the reduction to 60 m² for two-person households by (Zimmermann 2018) mentioned above, to a 20 % decrease in per capita floor area by 2050 (-17.5 Mt $CO_2$-eq./a) (Pauliuk and Heeren 2021) to zero net addition of floor area by 2030 (-7.6 Mt $CO_2$-eq./a) (Thomas et al. 2019). Measures targeting small groups of persons, such as house divisions with 0.5–1 % participation rates, achieve reductions as low as -0.04 Mt $CO_2$-eq./a (Fischer et al. 2020). Due to higher penetration rates, long-term time horizons generally yield larger reduction potentials.

### 3.2.2. Mobility sector

The mobility sector includes a multitude of GHG saving potentials (40 measures), see Figure 5. Most of them can be ascribed to the higher-level cluster *motorised individual transport*. Five of the measures are calculated via consumption-based accounting and are valid for the target years 2045 or 2050. The three highest saving potentials (*ride sharing*: 15.8 Mt $CO_2$-eq./a, *car sharing* 15.2 Mt $CO_2$-eq./a and *public transport* 14 Mt $CO_2$-eq./a) stem from the same LCA study (Prakash et al. 2023) in which all emissions from resource extraction, especially for electric vehicle batteries, to recycling are considered. The other LCA study (Pauliuk and Heeren 2021) also provides potentials on ride and car sharing, but they are significantly lower (3.4 Mt $CO_2$-eq./a and 1.2 Mt $CO_2$-eq./a). The most likely reason for this discrepancy is that this study only examines a selection of materials like steel, aluminium and plastics.

The different saving potentials relating to the *speed limit* can be explained by the different maximum speeds considered in the studies for different road types. For the mobility sector, realised and expected potentials are available, see Figure 10 in the Appendix. The realised one is the saving potential for *remote work* during the COVID-19 pandemic (Kreye et al. (2022); saving: 3.7 Mt $CO_2$-eq./a). The difference to the very low potential from NPM (2019) can most likely be explained by the time at which the study was written or published, i. e. before or after the lockdowns when remote work became the new normal for many people. The highest remote work potential (9.6 Mt $CO_2$-eq./a) is a theoretical one in Kreye et al. (2022), which assumes five more home office days compared to pre-Covid times for a 35 % remote work quota.

The discrepancy within the *reduce flights* cluster is due to the different ambition level of the measures: in the one with a higher saving potential, private air travel is halved (Fischer et al. 2016); in the other measure, inter-European short-haul flights are replaced when a rail trip to the same destination is no longer than four hours (Gaude et al. 2021).

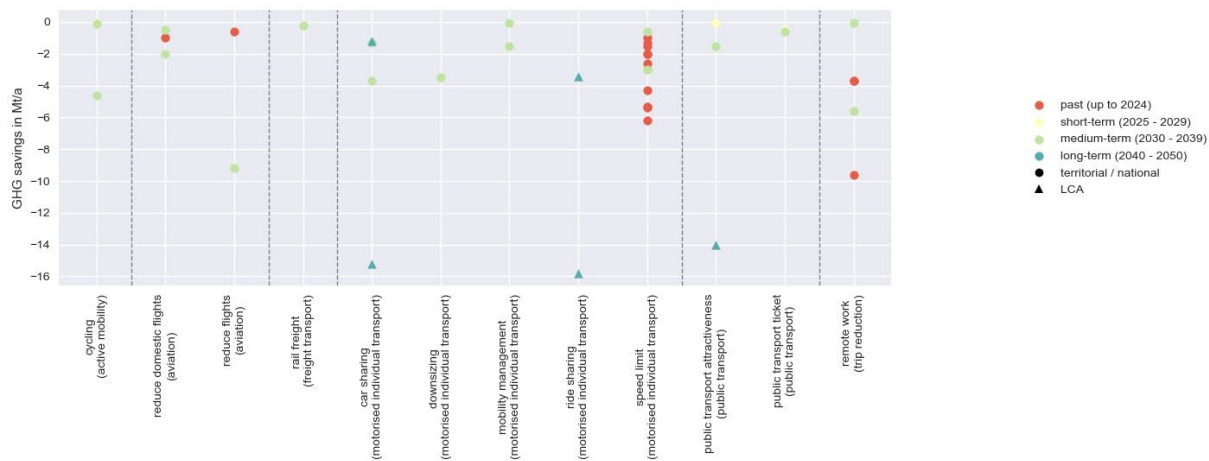

**Fig. 5** GHG saving potentials in the mobility sector by calculation method.

All 17 energy saving potentials in this sector are theoretical ones based on a territorial calculation approach, see Figure 11 in the Appendix. They cover four higher-level clusters - the same as for the GHG savings but without *freight transport* and *public transport*. The highest saving potential is identified for *remote work* in a study published in 2016 (Fischer et al. 2016), which anticipated very ambitious remote working modes with 40 % fewer business trips. The saving potential of 41.1 TWh/a is equivalent to nearly 6 % of the current energy consumption of the sector.

### 3.2.3.    Industry / production sector

Energy savings in the industry / production sector range between 0.5 and 31 TWh (Eerma et al. 2022). Five measures use a territorial approach, while two employ LCA methods, see Figure 12. Quantifications are either retrospective or for long-term horizons. Smaller and realised potentials are identified within the *circular economy* cluster, such as passenger spare part reuse, see Figure 6 (because the potentials are so close to each other, the four markings cannot be clearly distinguished). Larger savings stem from theoretical potentials on sharing economy initiatives (all products) and reducing production volumes of energy-intensive products (cluster material substitution).

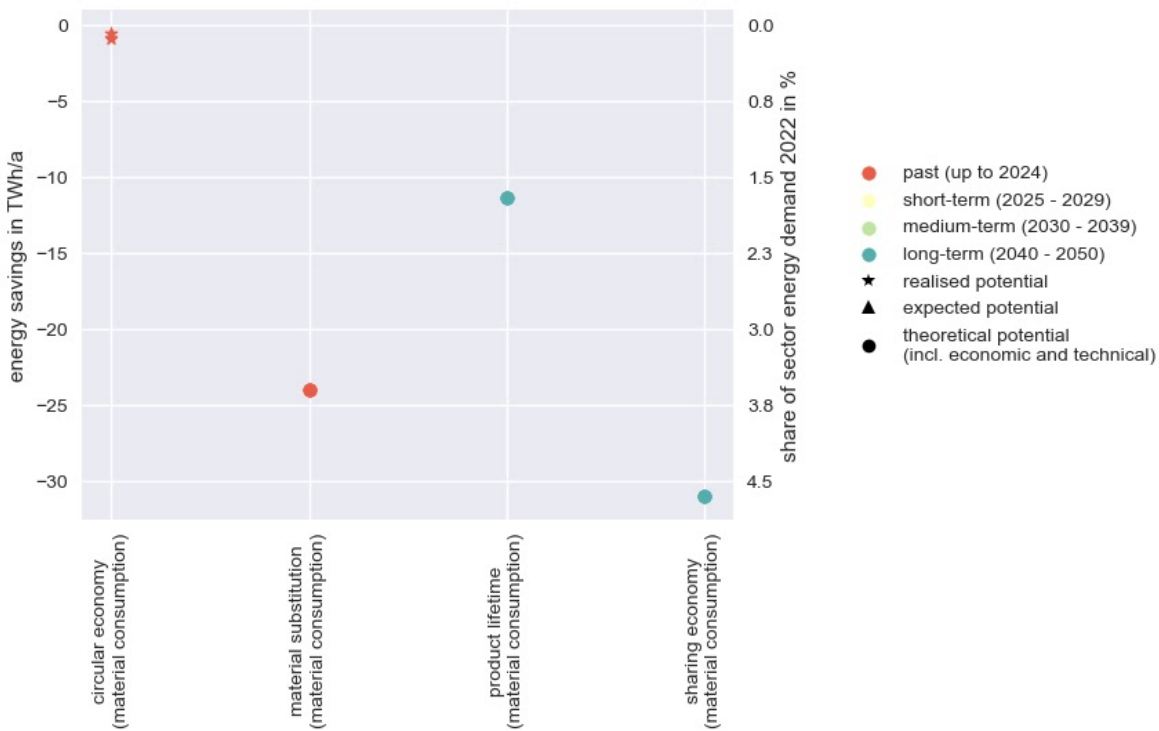

**Fig. 6** Energy saving potentials in the industry / production sector by type of potential. Final energy demand data was used from AGEB (2024).

The database identifies 15 measures, ten measures within the *product lifetime* cluster for GHG savings and one to two per other clusters, see Figure 13 in the Appendix. Except for one territorial quantification, all use an LCA approach and comprise theoretical potentials.

GHG reduction potentials vary widely, from 0.02–5.5 Mt $CO_2$eq./a for measures targeting specific product groups (e. g. lamps, furniture, household appliances) to 37.3 Mt $CO_2$eq./a for general product lifetime extensions (Zimmermann 2018). As with energy reductions, smaller reductions are associated with targeted product groups, while the highest potentials stem from broader lifetime extension measures across all products.

### 3.2.4.    Land-use and food sector

The database includes 32 entries of GHG saving measures in the land-use and food sector. Energy saving does not play a role in this sector and all entries are theoretical potentials. In terms of methodology, a few studies are based on LCA; the majority, however, uses a territorial approach, see Figure 7.

The potentials are grouped into three higher-level clusters. Most of the potentials can be identified for *natural carbon sequestration* and *diet changes*. Examples of potentials within the *natural carbon sequestration* cluster are a reduction of timber extraction from forests (68 Mt $CO_2$-eq./a, Reise et al. 2024) and the rewetting of agricultural land on drained peat soils (22 Mt $CO_2$-eq./a, Purr et al. 2021).

For the *diet change* cluster, the entry with the highest saving potential (28.7 Mt $CO_2$-eq./a, Prakash et al. 2023) is the most ambitious assumed diet change, with the Planetary Health Diet (PHD) realised by 2045.

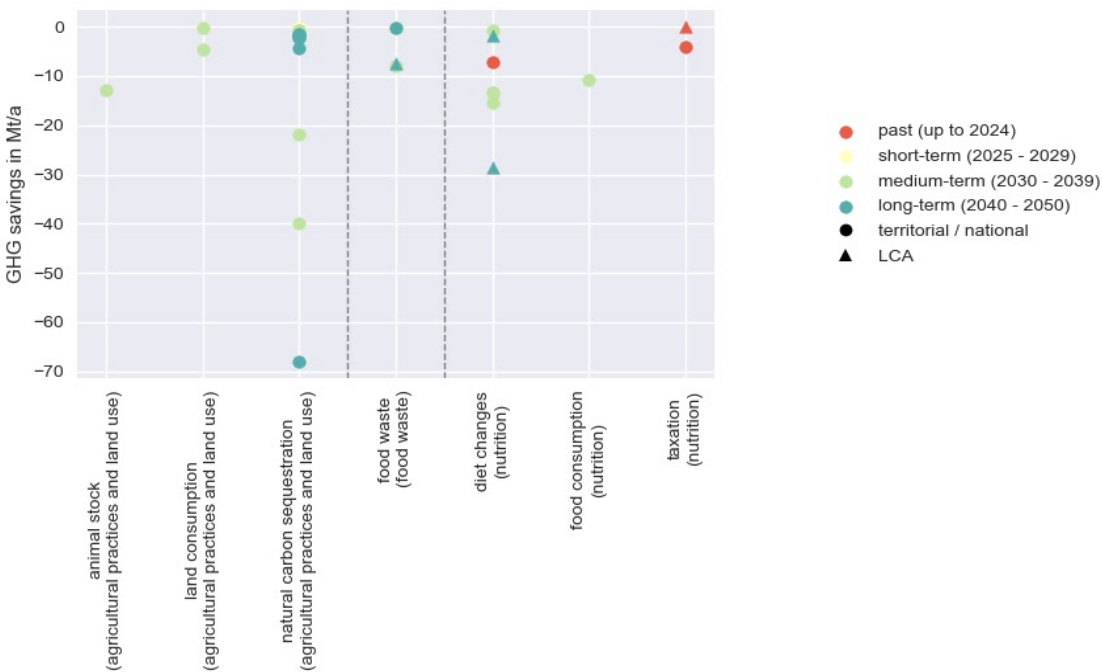

**Fig. 7** GHG saving potentials in the land-use and food sector by calculation method Characteristics of all potentials include type: theoretical.

## 3.3.  Gap analysis

As described in the methods section (chapter 2.4), we compared the entries of the potential database (wide scope) with the *European Sufficiency Policy Database* (Zell-Ziegler et al. 2025) to identify whether there are gaps in sufficiency quantification which future research can close. As the policy database only lists policy instruments, and the potential database includes all kinds of measures (66 % of them not directly linked to a policy instrument), the comparison at this level of detail is somewhat limited. Yet, the policy database includes policy targets and measures or actions that were compared to the clusters of the potential database.

As described above, in the *Potential Database*, most of the entries relate to the building sector. Many of these are behavioural changes, such as downsizing appliances or reducing heating, and there is no mention of whether they were voluntary or driven by policy instruments. The same is true for the entries on reduced living space, although there are many instruments to promote a reduction in living space, as shown in the *Policy Database*. Thus, in future assessments, the potentials could be better linked to policy instruments. With the exception of the measure or action category *vertical densification* of the *Policy Database*, we found a wide range of possible sufficiency measures quantified in this sector.

Most *Policy Database* entries focus on the mobility sector and are highly detailed. The quantified entries in the *Potential Database* lack this level of granularity, particularly for public transport. Future quantifications could disaggregate the savings further and explore savings from measures like *reducing parking*, *car-free city centres*, *city tolls* and *walking*. Currently, the potentials emphasise *speed limits* and *cycling*.

In the industry and production sector, we identified many potentials in the area of circular economy, which is a very broad field. Potentials were lacking on *product labelling* and *regional economy or local production*.

In the land-use and food sector, we are also lacking savings from the local production of food and feed, and there is a gap in the saving potential from reducing fertiliser use. Similar to measures in the building sector, the potential for dietary changes is often assumed without concrete policy instruments for changing eating habits. However, when pointing out gaps in this area, it should be noted that the land use and food sector is a long way from energy sufficiency and that many potentials and literature sources may not have been identified by our query.

Since cross-sectoral and more systemic sufficiency potentials were not the focus of this literature review, it is not surprising that the share of cross-sectoral policy instruments with corresponding quantification is relatively low. However, systemic cross-sectoral policies are very important and effective (Babiker et al. 2022) and should thus also be examined further.

## 3.4. Discussion of results

Through the literature review, we have identified the existing saving potentials for different sectors and clusters scattered throughout the literature. We then systematised these findings in one place and made them comparable by harmonising the units. However, despite this harmonisation, comparing the different saving potentials remains challenging, even for the same measure (e. g. speed limit or dietary change). This is because the potentials are heterogeneous in terms of factors such as time horizon, calculation method, type of potential, and relation to a baseline or reference value.

It became clear to the authors that the database needs a comprehensive set of metadata for each saving potential. Such metadata is essential for accurately presenting and comparing potentials as it highlights the key differences between them. The database created attempts to capture as much of the relevant metadata as possible in the different defined columns (see chapter 2.2) and additional commentary columns. However, the database can only be as precise as the primary sources are. Yet, there are much more relevant assumptions made in the studies that could not be collected in the database because they are, in most cases, not explicitly described in the studies. These assumptions either relate to (political) context and (normative) background or are underlying assumptions that influenced the resulting saving potentials. There is a need for more transparency and metadata to understand the assumptions behind quantified saving potentials.

We found that all potentials based on theoretical assumptions are higher than realised or expected potentials within a cluster, which is plausible. Potentials resulting from a consumption-based approach tend to be higher than those from a territorial approach, which is also plausible. Moreover, we found that for all sectors, the temporal scale was related to the level of savings to a minor extent.

Very robust saving potentials result from ex-post evaluations but the share of this data is low in the database (4 %). This is not surprising because sufficiency does not play a large role in current politics (Zell-Ziegler et al. 2021) and, as a result, there are not many implemented measures that can be evaluated. This underscores the importance of evaluating existing sufficiency projects and calculating their saving potentials.

The potentials also greatly differ in terms of the scope they cover, e. g. all household appliances in all German households or just fridges in households that include people over 60 years of age. The often narrow perspective may explain why some of the potentials are very low. For example, Fischer et al. (2020) examined very specific potentials for small target groups such as 'young couples starting a family' or 'older households in [their] own home'. Some potentials are only given for the local level like the municipal cycling projects in Paar et al. (2023) and some are only valid for very specific product groups like 'passenger car spare parts' in McKenna et al. (2013). These results show that sufficiency interventions often happen at small scales and can be quantified at this level of detail. This leads to the conclusion that sufficiency measures need to be scaled-up by businesses, politics or other actors and combined with each other or efficiency and consistency measures to realise their full potential.

Despite the described differences influencing the calculated saving potentials, the database provides insights into potential savings from sufficiency measures, guiding discussion on mitigation options beyond renewable energy deployment and energy efficiency. In this regard, our study links with others like the comparison of energy service demand indicators of different sufficiency scenarios by Wiese et al. (2024b) – the data presented in this study is also very useful to energy modellers and can, if applied in scenarios, result in energy and GHG saving potentials. The more saving potentials are added to the database in the future, the more robust insights regarding the potentials will become. This *Sufficiency Potential Database* can then better inform decision-makers and serve as a stronger foundation for energy modellers.

The systematic review showed that it is very difficult to catch relevant literature on quantified energy sufficiency measures via a quite simple search string. Most of the literature sources that led to entries in the *Sufficiency Potential Database* were known by the authors or their networks or added through snowball sampling. This is due to the fact that the term sufficiency is often not explicitly used in the literature sources. Rather, it is paraphrased as, for example, 'efficient use of living space' or is not necessary when talking about some sufficiency measures, e. g. for the promotion of public transport. To make the potential saving of energy sufficiency visible, it would be beneficial if common language would be used. If not, it remains a small-scale task to identify the relevant literature sources. Another issue is that almost three quarters of the literature sources used for the database is grey literature and thus it is unclear how the quality was ensured. There can be different reasons for this. Energy sufficiency measures are not yet widely implemented, leading to a lack of data for peer-reviewed studies. Additionally, the scientific community has not yet focused heavily on sufficiency as a mitigation option. Furthermore, studies on saving potentials are often commissioned by administrations or NGOs, for whom scientific publications may not be a priority.

Overall, it would be beneficial to include different communities and networks in the future expansion of the database to incorporate many more potentials from different fields. The current database provides a starting point for this.

# 4.  Conclusions

We have established a novel, openly available database with sufficiency saving potentials for GHG emissions and energy consumption for Germany. The literature review which fed the database led to 47 relevant sources from which we extracted 303 unique saving potentials. This demonstrates that quantified saving potentials are available in the literature and can be used by modellers and policymakers. We also believe that there are many more quantified saving potentials in the literature that were not identified through our query and expert networks. The database makes access, comparison and the use of data easy, particularly because it provides a conversion of all possible potentials to the same unit and a great amount of metadata.

Most quantifications were identified for the building sector (41 %), especially in the *appliances* cluster. The highest saving potential identified in the literature was for lowering the per capita living space to 30 $m^2$/person (150 TWh/a) and the second highest was for a combination of car sharing and public transport use (104 TWh/a). Both saving potentials are equivalent to 15 % of the current respective sectoral final energy demand. We thus confirm the high saving potentials stated in other review studies like IPCC (2022). Regarding GHG savings, the highest potential given in the literature was for a combination of reducing room temperature and optimising space use (118 Mt $CO_2$-eq./a). The reduction of living space / the effective use of space has thus the highest energy and GHG reduction potential within our database.

The database structure highlights the key metadata that enables saving potential data to be used further. These are especially assumptions on the reference scenario, if applicable, on the scope of calculation (territorial or consumption-based accounting) and on the type of the potential (theoretical - including technical and economic ones, expected or realised). Moreover, the saving measure itself (for sufficiency mainly the change of activity levels) needs to be described in detail. More open and transparent data provision or a standard for metadata on studies including saving potentials would be helpful[5].

The database is intended to be extendable or to serve as a blueprint. This means that more saving potentials for Germany and for other countries can be added or a similar database in other geographical contexts can be set up. An extension will make it more useful and lead to more robust potentials for the clusters. A starting point for modellers and researchers can be to fill the gaps identified: more disaggregated potentials for the mobility sector and more potentials on regional economies. To equalise the imbalance of the current database, it would be useful to have more studies that use a consumption-based approach (LCA method) and more bottom-up calculations that can also be used on local and regional levels, which often want to or could implement sufficiency. Backing the quantified measures by concrete policy instruments seems favourable as well. Furthermore, we need more evidence from ex-post evaluations of implemented sufficiency measures and a better link of GHG and energy saving potentials with other potentials such as job creation, biodiversity conservation, land use etc.

In summary, we expect the database to be useful for researchers and modellers as input for energy and climate scenarios and as a reference to which they could add new findings. For policymakers, we expect the database to help with identifying the potentials of energy sufficiency, design policy instruments to realise the potential and implement sufficiency as part of the energy transition and decarbonisation.

---

[5] Discussions, ideas and efforts in this direction can be found in these publications: Hülk et al. 2018; Speck et al. 2023; Dioha et al. 2023; Grether et al. 2024

# Appendix

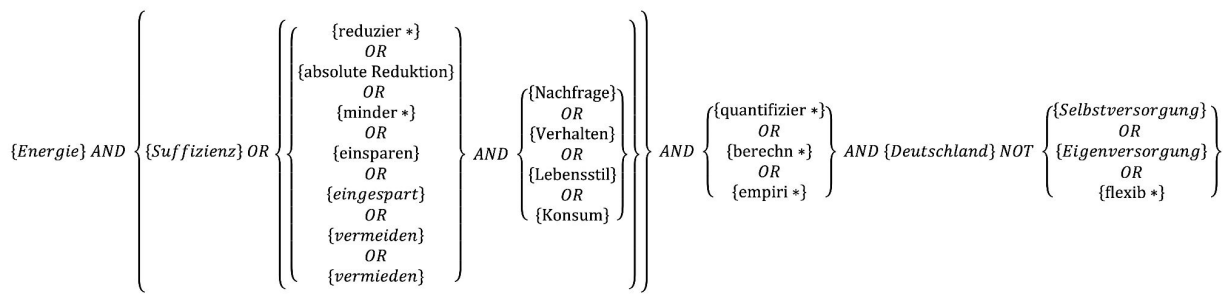

**Fig. 8** Search query for the German search

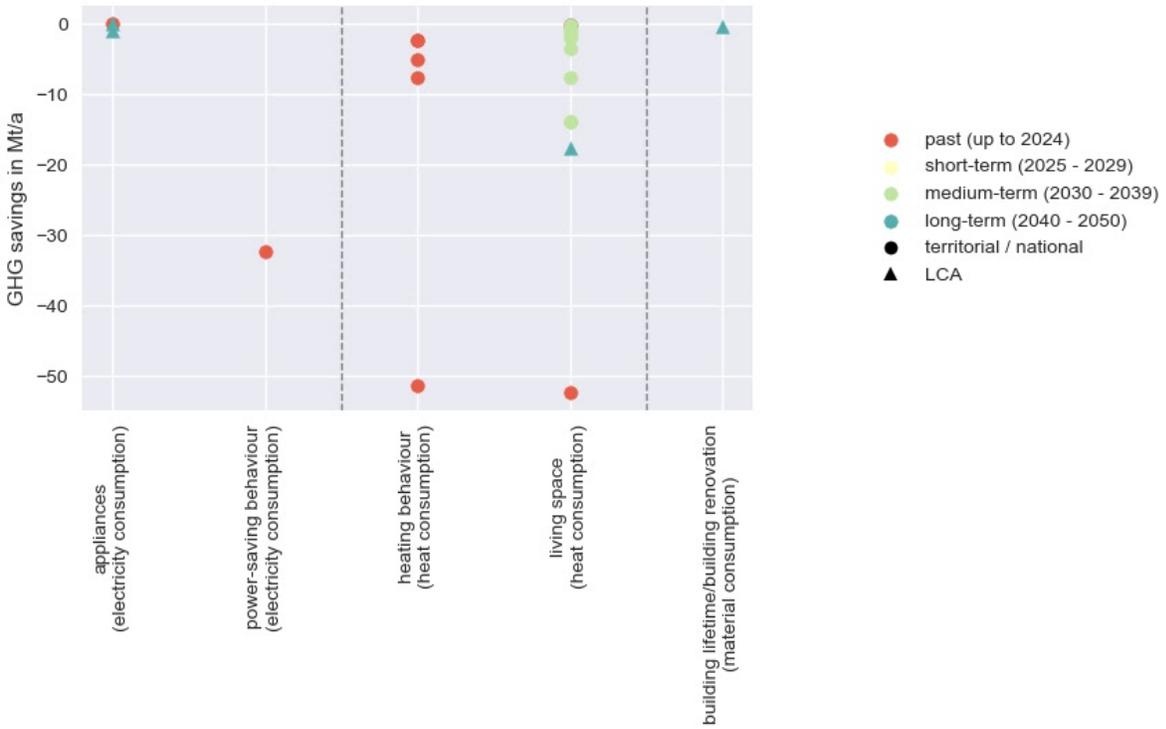

**Fig. 9** GHG saving potentials in the buildings sector by calculation method. Characteristics of all potentials include type: theoretical.

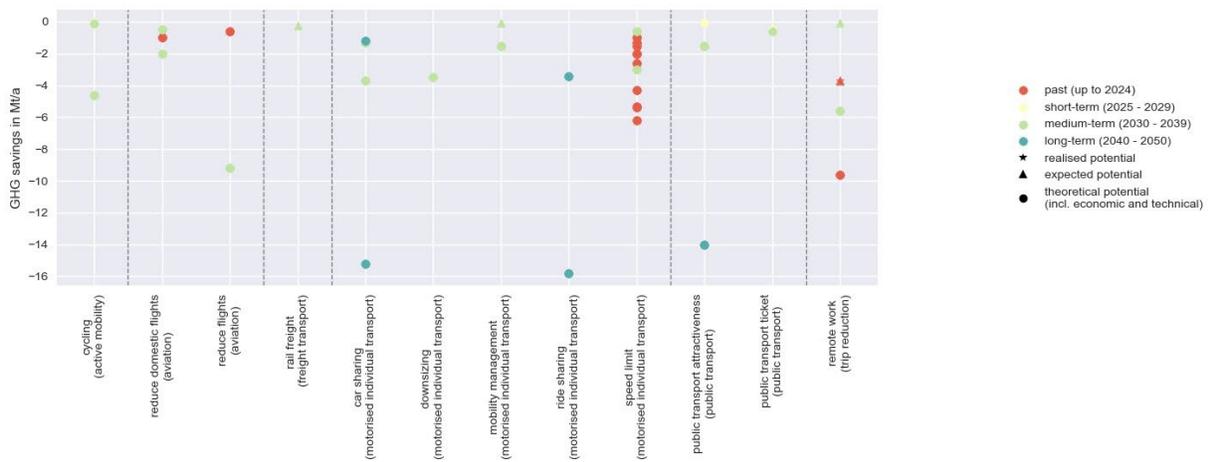

**Fig. 10** GHG saving potentials in the mobility sector by type of potential

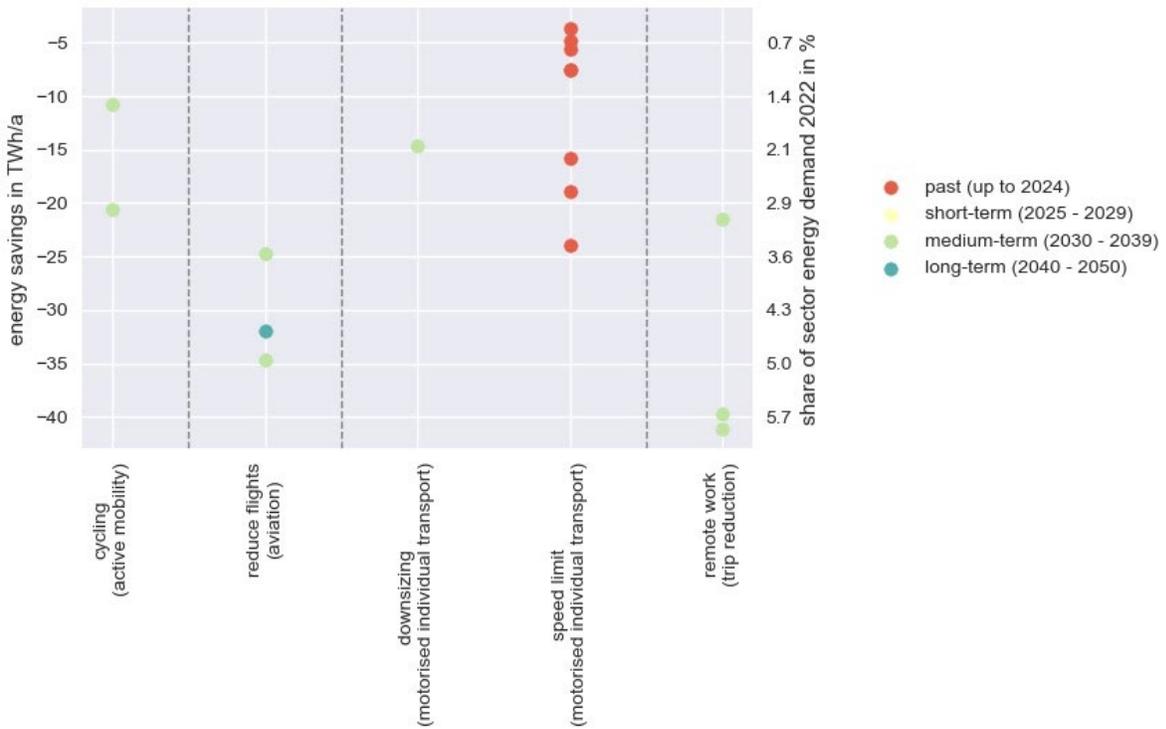

**Fig. 11** Energy saving potentials in the mobility sector. Characteristics of all potentials include territorial calculation method and type: theoretical. Final energy demand data for 2022 was taken from AGEB (AGEB 2024).

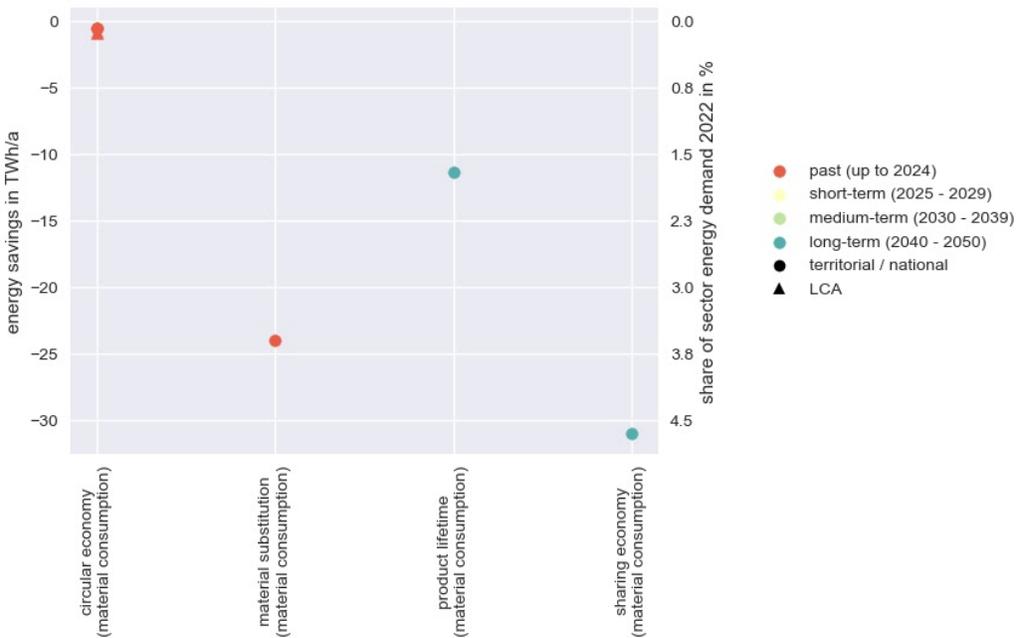

**Fig. 12** Energy saving potentials in the industry / production sector. Final energy demand data for 2022 was taken from AGEB (AGEB 2024).

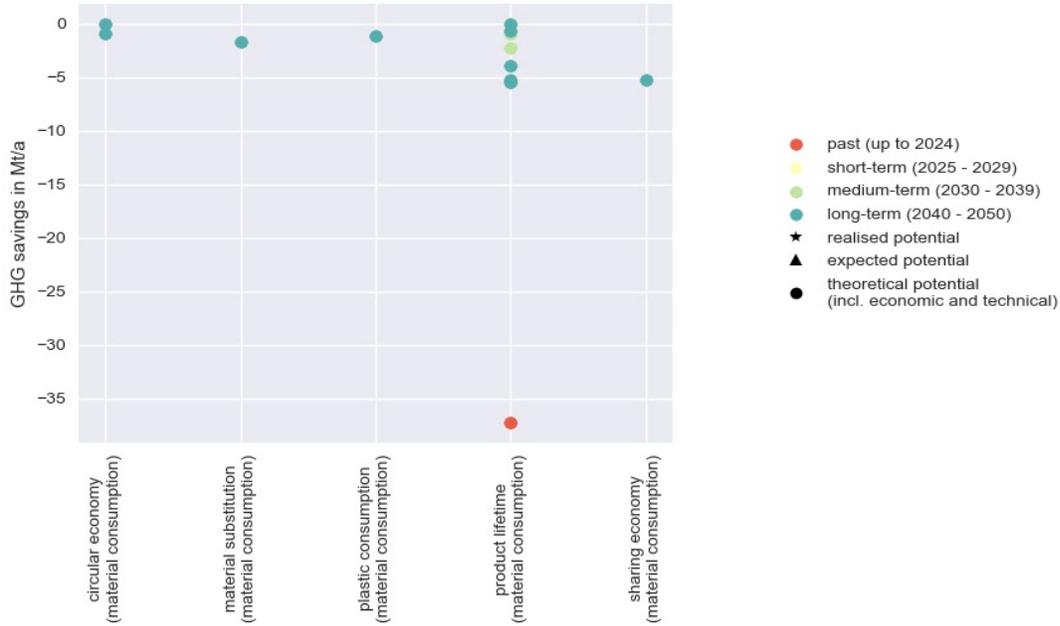

**Fig. 13** GHG saving potentials in the industry / production sector by type of potential

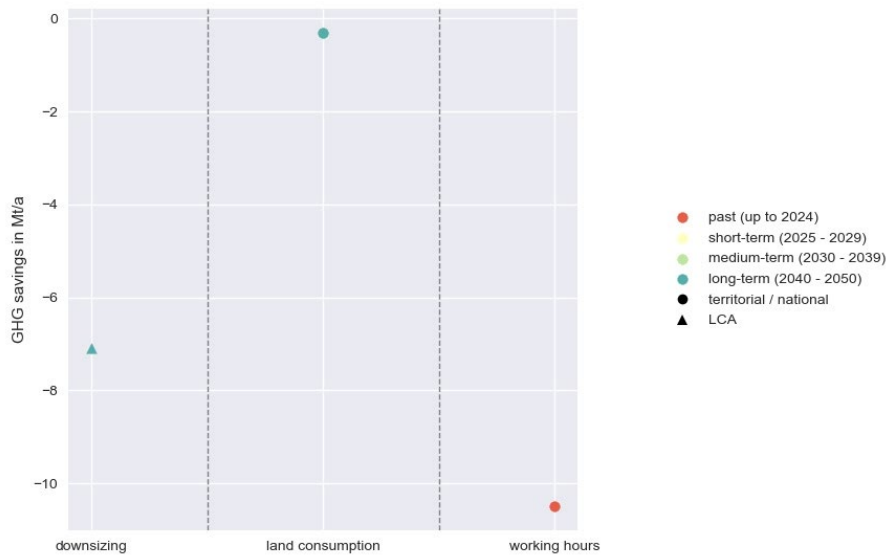

**Fig. 14** Cross-sectoral GHG saving potentials by type of potential. Characteristics of all potentials include territorial calculation method and type: theoretical.

# Supplementary Information: Sufficiency Potential Database for Germany (xlsx.-file)

The database is also available at https://doi.org/10.5281/zenodo.14779100

**Acknowledgements:**

This work was inspired by the Energy Demand Changes Induced by Technological and Social Innovations (EDITS) network, an initiative coordinated by the Research Institute of Innovative Technology for the Earth (RITE) and the International Institute for Applied Systems Analysis (IIASA) and funded by the Ministry of Economy, Trade and Industry (METI), Japan.

We thank Prof. Dr. Stefan Heiland, Technische Universität Berlin, for fruitful discussions and helpful feedback on the concept, methods and results of this paper. We thank Dr. Vanessa Cook, Oeko-Institut Berlin, for her excellent proof-reading.

**Funding:**

This research was conducted as part of the research project 'The role of energy sufficiency in energy transition and society' funded by the German Federal Ministry of Research, Technology and Space (BMFTR), within the framework of the Strategy Research for Sustainability (FONA), as part of its Social-Ecological Research funding priority [grant numbers 01UU2004C (CZZ and KD), 01UU2004B (DS) and 01UU2004A (FW)]. CB acknowledges funding from the doctoral scholarship programme of Deutsche Bundesstiftung Umwelt (DBU). MS would like to thank the German Federal Government, the German State Governments, and the Joint Science Conference (GWK) for their funding and support as part of the "NFDI4Energy" consortium. The work was (partially) funded by the German Research Foundation (DFG)—501865131 within the German National Research Data Infrastructure.

Responsibility for the content of this publication lies with the authors.